\documentclass[fleqn,twocolumn, superscriptaddress]{revtex4}
\usepackage{epsfig}
\usepackage{graphicx}
\usepackage{color} 
\usepackage{amsfonts}
\usepackage{amssymb}

\newcommand{\ct}{\cite}
\newcommand{\bi}{\bibitem}
\newcommand{\be}{\begin{equation}}
\newcommand{\ee}{\end{equation}}
\newcommand{\ba}{\begin{eqnarray}}
\newcommand{\ea}{\end{eqnarray}}

\newcommand{\non}{\nonumber}
\newcommand{\bra}[1]{\langle #1|}
\newcommand{\ket}[1]{|#1\rangle}

\begin{document}
\title{Local shortcut to adiabaticity for quantum many-body systems}

\author{Victor Mukherjee}
\affiliation{Department of Chemical Physics, Weizmann Institute of Science, Rehovot 76100, Israel}
\affiliation{Institute for complex quantum systems $\&$ Center for Integrated Quantum
	Science and Technologies (IQST), Universit\"at Ulm, D-89069 Ulm, Germany}

\author{Simone Montangero}
\affiliation{Institute for complex quantum systems $\&$ Center for Integrated Quantum
	Science and Technologies (IQST), Universit\"at Ulm, D-89069 Ulm, Germany}

\author{Rosario Fazio}
\affiliation{ICTP, Strada Costiera 11, I-34151 Trieste, Italy}
\affiliation{NEST, Scuola Normale Superiore $\&$ Istituto Nanoscienze CNR, I-56126 Pisa, Italy}

\begin{abstract}
We study the environment assisted local transitionless dynamics in closed spin systems driven through quantum critical points. In general shortcut to adaiabaticity (STA) in quantum critical systems requires highly non-local control Hamiltonians. 
In this work we develop an approach
to achieve local shortcuts to adiabaticity (LSTA) in spin chains, using local control fields which scale polynomially with the system size, following universal critical exponents. We relate the control fields to reduced fidelity susceptibility and use transverse Ising model in one dimension to exemplify our generic results.  We also extend our analysis to 
achieve LSTA in central spin models.
\end{abstract}

\maketitle
\section{Introduction}
The dynamics of closed quantum many-body systems is currently very intensively investigated as it is at the heart of the understanding of many
fundamental problems in modern science~\cite{polkovnikov,eisert} as well central for emerging quantum technologies~\cite{georgescu}. In several 
important cases it is of great importance to be able to have means to control the dynamics of these complex systems. This is for example 
the case of the  preparation of certain desired many-body states. To this aim optimal control techniques have been specially devised for many-body 
systems~\cite{doria,lloyd} and successfully implemented experimentally~\cite{rosi,vanfrank,vanfrank1}. Optimal control of many-body systems have proven 
to be helpful also in reducing the formations of defect on traversing a critical point~\cite{caneva} as well as  in cooling~\cite{rahmani}, just to mention
some examples.

Among the various way to engineer the  dynamics of a quantum system the so called shortcuts to adiabaticy (see~\ct{torrontegui13}  for 
a review of the field) are attracting an increasing attention. By judicious choice of time-dependent contributions to the Hamiltonian, shortcuts 
to adiabaticity (STA) allows for a truly adiabatic evolution at a finite speed.  A quantum system prepared initially in a given eigenstate of an Hamiltonian 
$H_{free}$ can remain in the corresponding instantaneous eigenstate of $H_{free}(t)$ on adding a specially designed term $H_c(t)$ to the Hamiltonian 
governing the evolution of the system~\ct{berry09, rice03}.  Developing different protocols for achieving STA has been the focus of several studies,
for example the application of counter-adiabatic terms~\ct{berry09, rice03, campo13},  the fast forward approach~\ct{masuda08} to name a few together 
with its application to two and three level atoms \ct{chen10} and to universal quantum computation \ct{santos15}. STA was also extended to the case of  non-Hermitian Hamiltonians~\ct{muga11}, 
and to open quantum  systems governed by a Lindblad dynamics~\ct{vedral14}. The theoretical efforts have been  complemented by experimental 
implementation of STA protocols~\ct{bason12, monroe13}.

Shortcut to adiabaticity was also extended to control the dynamics of many-body system governed by the transverse Ising~\ct{zurek12} and 
Lipkin-Meshkov-Glick~\ct{campbell} Hamiltonians. A distinct feature emerging from these works is the increasing complexity of the Hamiltonian 
$H_c$ as a result of gaps closing in the spectrum. Suppose to apply STA to a system initially prepared in the ground state of a spin system. 
The correction $H_c$ to be implemented will contain multi-spin interaction which become progressively important when $H_{free}$ is close to a 
quantum critical point (QCP); a QCP is associated with closing of the energy gap between the ground state and the first excited state, resulting in diverging correlation length and time, 
as well as onset of spontaneous
 symmetry breaking ~\ct{sachdev99, zurek12}. This effect is intimately related to the so called Kibble-Zurek mechanism~\ct{zurek05, polkovnikov05}.
In general, a system starting from an initial ground state evolves adiabatically and always remains in its instantaneous ground state, when it is 
in presence of a Hamiltonian changing sufficiently slowly in time. However,  when a parameter of the Hamiltonian is changed across its critical 
value at an arbitrary rate $\tau^{-1}$, as explained above, the energy gap vanishes. This results in breakdown of 
the adiabatic theorem of quantum mechanics near the the critical point thus invariably resulting in non-adiabatic excitations, which  
vanish only in the ideal, and experimentally unachievable, limit of $\tau \to \infty$ \ct{zurek05, polkovnikov05}. It is possible to cross a critical point 
without defect generation either by optimal control~\cite{caneva,vanfrank1} or by means of STA~\cite{zurek12,campbell}. In both cases, in different ways, 
the existence of the critical point manifest in the need of an increasing complex control protocol.

So far STA was defined as to have the system  in the instantaneous eigenstate with unit fidelity. For many-body system it is however meaningful 
to relax this requirement and construct a LSTA protocol so to have strict adiabaticity guaranteed only in a subsystems of an 
extended system. The LSTA might be applied while manipulating qubits trapped in N-V centres or atoms trapped in optical lattices where, e.g., one might be interested in preparing 
only the center of the trap with less excitation as possible \ct{steiner10, gruber97, anderson11}.

\begin{figure}[ht]
\begin{center}
\includegraphics[width=0.95\columnwidth,angle = 0]{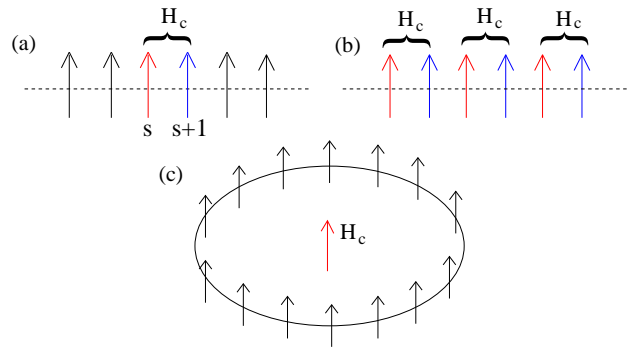}
\end{center}
\caption {(Color online) Schematic diagram showing LSTA applied to a (a) single qubit, (b) half of the spin system and (c) in a central spin model.
In (a) and (b) the control Hamiltonian $H_c(t)$ consists of a local field acting on the red spin under consideration and an interaction with a 
neighboring blue spin, whereas LSTA is achieved by only a local field acting on the central spin in (c).}
\label{schem1}
\end{figure}

The question can be formulated as follows. {\it The goal is to find a correction term $H_c$ such that the all the local observables (defined within a 
given length $L_{STA}$ of the total system) coincide with those calculated with the instantaneous ground state. The state does not need to be 
close to the eigenstate, only the reduced density matrices (in the region $L_{STA}$) should coincide.}
In this work we deal with LSTA by application of local control Hamiltonians only. Our analysis aims at engineering environment 
assisted adiabatic dynamics in subsystems, whose size can range from a single qubit to half of the total system,
by application of local control Hamiltonians. To this aim we consider a generic spin system driven through a QCP
at a finite rate. We achieve LSTA in a single spin (shown by spin $s$ in red in Fig. (\ref{schem1}a))  by application of control fields acting on this spin
under consideration, in addition to an interaction with a nearest neighbor auxiliary spin (shown by spin $s+1$ in blue in Fig. (\ref{schem1}a)). Further, we present 
the possibility of applying the same control Hamiltonian $H_c(t)$ at multiple sites to achieve STA in up to half of the spin system (Fig. (\ref{schem1}b)). 
Our analysis points towards generic polynomial scaling relations followed by the control fields, which again can be related to reduced fidelity susceptibility
of the system.  We have also extended our analysis to LSTA in quantum critical Hamiltonians which include long range interaction terms. In particular, 
we focus on a central spin model where the central spin is coupled uniformly to all spins of its  environmental spin chain (see Fig. (\ref{schem1}c)). 
Analogous to the previous case, we develop mechanisms to achieve LSTA using local fields which follow generic scaling relations with the size of the 
environment. 

The paper is organized as follows: We focus on local STA in a free Fermionic spin model with nearest neighbor interactions in section \ref{STA1A}; here we present control schemes and derive
scaling relations followed 
by the control fields. In section \ref{STA1B}, we verify  our generic results using the specific case of a one dimensional spin $1/2$ Ising 
model in presence of a transverse field and show that LSTA in this case is possible by application of local fields which scale logarithmically with the system size.
We extend our analysis of local STA to spin systems
subjected to long-range interactions in section \ref{STA2A}, where we focus on transitionless driving in a generic central spin model. Once more, we exemplify our results using a transverse 
Ising spin chain environment in section \ref{STA2B}. Finally we conclude in section \ref{concl}.

\section{Local transitionless dynamics in a spin system with nearest neighbors interactions}

\subsection{Free-Fermionic model}
\label{STA1A}
We first consider a $d$ dimensional translationally invariant time dependent spin Hamiltonian of the form
\ba
H_{free}(t) = \sum_{\bf{j}, \alpha} \left[\mathcal{A}^{\alpha}_{free}(t) \sigma_{\bf j}^{\alpha}  \sigma_{\bf{j} + \bf{r}}^{\alpha}  + \mathcal{B}^{\alpha}_{free}(t) \sigma_{\bf j}^{\alpha}\right].
\label{hamilspin}
\ea
Here $\mathcal{A}^{\alpha}_{free}$ denotes the interaction strength along the spatial dimension $\alpha$ ($= x, y, z$)
between nearest neighbor spins at $\bf{j}, \bf{j} + \bf{r}$, while
$\mathcal{B}^{\alpha}_{free}$ is
the external magnetic field along $\alpha$ axis; $\sigma_{\bf j}^{\alpha}$ are the corresponding Pauli spin matrices. We assume our spin system (\ref{hamilspin}) can be mapped to a free-Fermionic
Hamiltonian of the form
\ba
H_{free}(t) &\equiv& \sum_{\bf k} C_{\bf k}^{\dagger} \mathcal{K}(t) C_{\bf k} \non\\ 
\mathcal{K}(t) &=& \left[ \begin{array}{cc} a_{\bf k}^z(\lambda(t)) & a_{\bf k}^x - i a_{\bf k}^y \\
a_{\bf k}^x + i a_{\bf k}^y & -a_{\bf k}^z(\lambda(t)) \end{array} \right].
\label{ff}
\ea
This is indeed true for a wide range of extensively studied models, including the Ising and the XY
models in $d = 1$, as well as the Kitaev model in $d = 1$ and $d = 2$, which can be mapped from spin to free-Fermionic representation by Jordan-Wigner transformation \ct{lieb61, dziarmaga09}.
 $C_{\bf k}^{\dagger} = (c_{{\bf k}1}^{\dagger}, c_{{\bf k}2}^{\dagger})$ are the 
Fermionic creation operators and the 
summation runs over all the independent ${\bf k}$ modes.
$\vec{a}_{{\bf k}}$ are functions of $\mathcal{A}^{\alpha}_{free}(t), \mathcal{B}^{\alpha}_{free}(t)$ and the momentum modes ${\bf k}$, with the functional form depending on the specific model under consideration. The
parameter $\lambda$, which is a function of $\mathcal{A}^{\alpha}_{free}$ and $\mathcal{B}^{\alpha}_{free}$ as well, denotes the distance from the QCP 
$\lambda = 0$, where the energy gap vanishes, i.e., $a_{\bf \hat{k}}^z(\lambda = 0) = a_{\bf \hat{k}}^x = a_{\bf \hat{k}}^y = 0$, $\hat{k}$ being the critical mode. Whenever it is possible, 
transforming back to the spin
representation (\ref{hamilspin}), one can get equivalent critical fields
$\mathcal{A}^{\alpha}_{free} = \hat{\mathcal{A}}^{\alpha}_{free}$ and $\mathcal{B}^{\alpha}_{free} = \hat{\mathcal{B}}^{\alpha}_{free}$, say, where the energy gap vanishes \ct{sachdev99}. 
We assume the spin system initially starts in its ground state at $t \to -\infty$. In the absence of any control the global 
Hamiltonian $H_{free}(t)$ drives the system through the QCP, thus resulting in excitations for any finite rate of driving.
This eventually changes the reduced single site density matrix $\rho_1(t)$ from its instantaneous ``ground state form'' $\mathcal{G}(t)$,  
the single site density matrix obtained for perfect adiabatic evolution of the whole spin system.

We consider the spin model with a local control Hamiltonian $H_c(t)$ of the form
\ba
H(t) &=& H_{free}(t) + H_c(t) \non\\
H_c(t) &=& h_z(t)\sigma_s^z - \Delta_y(t) \sigma_s^y + J(t)\sigma_s^z \sigma_{s+1}^z,
\label{hcon}
\ea
applied in addition to the original free Fermionic Hamiltonian $H_{free}(t)$ near the QCP. The control Hamiltonian $H_c(t)$ is chosen so as to impose the LSTA condition 
$\rho_1(t) = \mathcal{G}(t)$ (see below), and can be considered as
equivalent to the counteradiabatic 
Hamiltonians for spin $1/2$ systems considered in related works on STA \ct{berry09, rice03, campo13}. In case of higher dimensional systems one can arrive at similar control Hamiltonians by generalization of 
Pauli matrices to operator basis in higher dimensions.
We note that the
dynamics is adiabatic for almost the entire length of time away from the QCP, when the relaxation time $\zeta_{\rm rel} \sim \lambda(t)^{-\nu z}$ of the system is much less than the time scale 
$\zeta_{H} = \lambda(t)/\dot{\lambda}(t)$ at which the Hamiltonian is changed, where $\nu$ and $z$ are respectively the correlation length and correlation time exponents near a QCP.
In this regime the rate of change of the Hamiltonian is slow enough for the system to follow its instantaneous ground state, and we have $\rho_1(t) \approx \mathcal{G}(t)$ even in absence 
of any control field. In contrast, the adiabaticity condition breaks down near the QCP $\lambda = 0$, where $\zeta_{\rm rel} \gg \zeta_{H}$ \ct{dziarmaga09, damski06, mukherjee07}. In this regime the system gets excited 
in absence of any control, thereby highlighting the importance of STA and LSTA near a QCP. In our scheme the
fields $h_z(t)$ and $\Delta_y(t)$ control
the phase and energy level populations of the spin respectively near the QCP,
whereas we control the purity of the spin $s$ by tuning its interaction strength $J(t)$ with a nearest neighbor spin $s+1$. 
One can expect the control fields (\ref{hcon}) to depend on the the single site reduced fidelity 
susceptibility $\chi_s$, which
quantifies the rate of change of $\mathcal{G}$ as a function of the system parameters. As shown in some related works,  $\chi_s$ as 
well as the response time of the system can be expected to diverge with 
the system size near 
the QCP \ct{ma08, you09, ma10}.

As we will show later, for small dimensions, the control field strength should diverge at the QCP, similar to the response time and $\chi_s$, given the diverging correlation length in the system.
Thus we can neglect the effects of $H_{free}(t)$ on the 
two spins under consideration in comparison to that of the diverging control fields near the QCP, and approximate the dynamics of
the two spin ($s, s+1$) density matrix $\rho_2$ by
\ba
\frac{\partial \rho_2 (t)}{\partial t} \approx -i\left[H_c(t),\rho_2(t)\right],
\label{vn}
\ea
where we have set $\hbar = 1$.

Finally, after using Eqs. (\ref{hcon}) and (\ref{vn}), tracing out the auxiliary spin
 and imposing the condition $\rho_{1}(t) = \mathcal{G}(t)$ near the QCP, one gets
\ba
\Delta_y (t) &=& \frac{1}{2\rm{Re}\left[\mathcal{G}^{1,2}\right]}\frac{\partial \mathcal{G}^{1,1}}{\partial t},
\label{delygen}
\ea
where $\mathcal{G}^{i,j}(t)$ denotes the $(i,j)$th element of $\mathcal{G}(t)$. Further, one has $\mathcal{G}^{1,1} = \left(1 + \langle\sigma_z\rangle_G \right)/2$, and
$\mathcal{G}^{1,2} =\langle\sigma^{-}\rangle_G$, where the single site expectation values are calculated in the ground 
state of the spin system. It is possible to obtain a generic scaling relation for $\Delta_y$ by restricting ourselves to the case 
\ba
\mathcal{G}^{1,1} &=& \frac{1}{2}\left(1 + \left< \sigma_z\right>_G \right) = \frac{1}{L^d}\sum_{\bf k} \mathcal{G}_{\bf k}^{1,1},
\label{spinF}
\ea
where $L$ is the length of the system and $\mathcal{G}_{\bf k}(t)$ denotes the  reduced density matrix of the ${\bf k}$th mode in the instantaneous ground state of $\mathcal{K}(t)$ (see eq. (\ref{ff})).
The above assumption (\ref{spinF}) is valid in translationally invariant spin chains, including the transverse Ising model and $XY$ model,
where the total number of up spins correspond to the total number of Fermions present in an equivalent free Fermionic picture. Following the above arguments, one finally arrives at the scaling forms
\ba
\Delta_y &\propto& \frac{\dot{\lambda}}{\rm{Re}\left[\mathcal{G}^{1,2} \right]} L^{z-d} ~~~~~~~\text{for}~ \lambda \ll L^{-1/\nu} \non\\
 &\propto& \frac{\dot{\lambda}}{\rm{Re}\left[\mathcal{G}^{1,2} \right]} \lambda^{-\nu(z-d)} ~~\text{for}~\lambda \gg L^{-1/\nu},
\label{scaly}
\ea
 (with $\nu z = 1$), and the constant of proportionality depends on the detailed form of $\vec{a}_{\bf{k}}$.
Note that $\Delta_y \sim \ln L$ at the QCP for $z = d$ and the approximation ($\ref{vn}$) might not be valid for $z<d$ when $\partial \mathcal{G}^{1,1}/\partial t$, and hence
$\Delta_y$, fail to diverge near the QCP.
In case of $\mathcal{G}^{1,2} = \left<\sigma^{-}\right>_G = 0$, which can arise due to symmetries in the model, we replace 
$\rm{Re}\left[\mathcal{G}^{1,2}\right]$ with a small infidelity parameter $\epsilon$ in Eq. (\ref{scaly}) in order to have a finite $\Delta_y$. Additionally, one can solve for $h_z(t)$ and $J(t)$ as 
well using Eq. (\ref{vn}), 
which needs detailed knowledge about the 
system. However, they can be expected
 to diverge with system size near the QCP as well; as shown below for the transverse field Ising model.

Furthermore, one may apply the same control Hamiltonian in multiple sites in order to achieve macroscopic 
adiabatic passage through QCP for up to $L/2$ spins located in the even or odd sublattice (assuming $L$ even). To this end we apply a simple additive control Hamiltonian of the form
\ba
H_{c,L}(t) = \sum_{j=0}^{L/2} \left[ h_z(t)\sigma_{2j}^z - \Delta_y(t) \sigma_{2j}^y + J(t)\sigma_{2j}^z \sigma_{2j+1}^z \right],
\label{hcon_global}
\ea
where we have assumed periodic boundary condition. As seen from Eq. (\ref{hcon_global}), the control Hamiltonian $H_{c,L}(t)$ couples all the even spins with their nearest neighbors and allows us to generate
local adiabatic passage through the QCP for half of the spin chain without the need of introducing highly non-local fields, 
even though the correlation length diverges near the QCP.  The above Eqs. (\ref{delygen}) - (\ref{hcon_global}) are the main results of our paper and they clearly show the possibility of engineering LSTA
through a quantum critical point
by application of local fields and local interactions only. 

Interestingly, as mentioned above one can arrive at an estimate of $\Delta_y$ from $\chi_s$. One has 
\ba
\chi_0 \sim \left(\frac{\partial \mathcal{G}^{1,1}}{\partial \lambda}\right)^2
\label{susc}
\ea
near a QCP, where we have  assumed $\partial \mathcal{G}^{j,j}/\partial t \gg \partial \mathcal{G}^{i,j}/\partial t$ 
($i \ne j$), which is indeed the 
case if $\left<\sigma^{\pm}\right>_G = 0$ \ct{ma08, ma10}. 
The above arguments finally give us
\ba
\Delta_y \propto \frac{\partial \mathcal{G}^{1,1}}{\partial t} = \dot{\lambda} \frac{\partial \mathcal{G}^{1,1}}{\partial \lambda} 
\sim \dot{\lambda}\chi_0^{1/2}.
\label{suscsc}
\ea

We note that following the technique introduced above, shortcut to adiabaticity in systems of arbitrary size, as compared to 
singe qubits considered here, might require an 
environment, formed by auxiliary spins, which is at least as large as the system 
itself, as suggested by a recent work on purification of open quantum systems \ct{viola14}. However, an exact analysis is
beyond the scope of our present work.

\subsection{Ising model in presence of a transverse field}
\label{STA1B}
Let us now elucidate the generic results obtained above using the exactly solvable example a spin $1/2$ transverse 
Ising model in one dimension, represented by the Hamiltonian
\ba
H_{free, I}(t) = -\sum_{j} \sigma_j^x \sigma_{j+1}^x - \mathcal{B}^z_{free}(t)\sum_j \sigma_j^z,
\label{hamil}
\ea
where we have taken the ferromagnetic interaction strength $\mathcal{A}^{x}_{free}$ to be unity (see Eq. (\ref{hamilspin})).
We assume $\mathcal{B}^z_{free} = t/\tau$, $-T < t < T$ for some arbitrary $T\gg\tau>0$. 
The above model (\ref{hamil}) has QCPs at $\mathcal{B}^z_{free} = \hat{\mathcal{B}}^{z}_{free} = \pm 1$, and 
can be decoupled into independent two level
systems characterized by their corresponding momentum $k$ modes  (as in Eq. (\ref{ff})) \ct{lieb61, pfeuty70, barouch71, bunder99}. We start with an initial spin up paramagnetic ground state at 
$t = -T$. The off-diagonal terms of the 
coarse grained single site density matrix  $\rho_1(t)$ vanish due to decoherence, while $\mathcal{G}^{1,2} = 0$ for all $t$ due to symmetry of the Hamiltonian (\ref{hamil}) \ct{levitov06}.
On the other hand,
one can show that in the absence of any control the  diagonal elements evolve very slowly with time far away from the QCP (when $|\mathcal{B}^{z}_{free}| \gg 1$), 
as well as near the QCP, as a
signature of critical slowing down.

\begin{figure}[ht]
\begin{center}
\includegraphics[width=0.95\columnwidth,angle = 0]{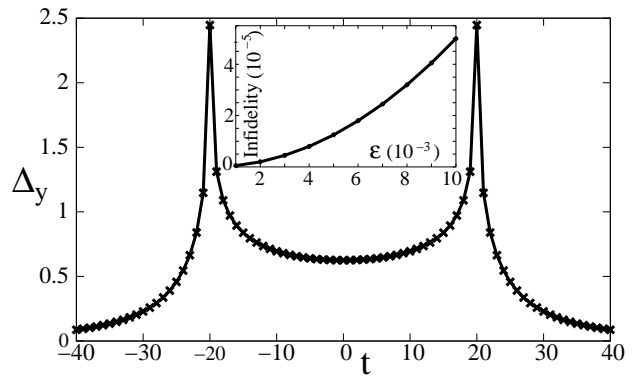}
\end{center}
\caption {Plot showing $\Delta_y$ as a function of time $t$  for $\tau = 20$, $L = 1000$ and $\epsilon = 0.01$. $\Delta_y$ rises sharply near the QCPs. Inset: Variation of infidelity between
$\mathcal{G}(t)$ and $\rho_1(t)$ as a function of $\epsilon$ at a QCP $t = -20$. All energies and inverse times are expressed in units of $\mathcal{A}^{x}_{free}$.}
\label{delta_vs_t}
\end{figure}

\begin{figure}[ht]
\begin{center}
\includegraphics[width=0.95\columnwidth,angle = 0]{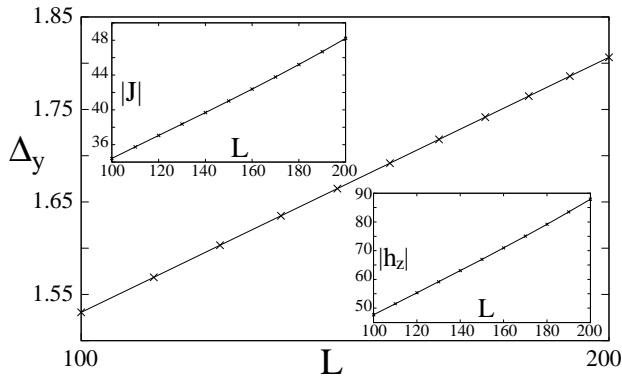}
\end{center}
\caption { Variation of $|\Delta_y|$ with system size $L$ at the QCP for $\tau = 20$, $t = -20$ and $\epsilon = 0.01$. $|\Delta_y|$ increases as $\rm{ln} L$.
Insets: Variation of $|J|$ and $|h_z|$  with system size $L$ at the QCP for $\tau = 20$, $t = -20$ and $\epsilon = 0.01$. All energies and 
inverse times  are expressed in units of $\mathcal{A}^{x}_{free}$.}
\label{delta_vs_L}
\end{figure}

\begin{figure}[ht]
\begin{center}
\includegraphics[width=0.95\columnwidth,angle = 0]{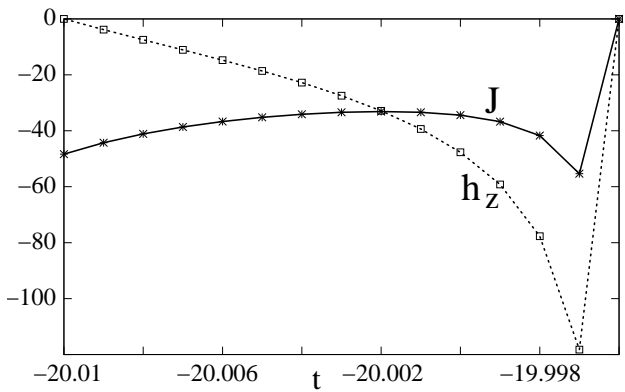}
\end{center}
\caption {Variation of $h_z$ and $J$ as a function of time near the QCP at $t = -20$ for $L = 100$, $\tau = 20$ and $\epsilon = 0.01$. 
All energies and inverse times  are expressed in units of $\mathcal{A}^{x}_{free}$.}
\label{hJ}
\end{figure}

In comparison, study of the single site ground state density matrix $\mathcal{G}$
shows
\ba
\frac{\partial \mathcal{G}^{1,1}}{\partial t} &\approx& \frac{1}{4\tau |\frac{t}{\tau}|^3} ~~~\text{for}~ |\frac{t}{\tau}| \gg 1  \non\\
&\approx& \frac{\ln L}{2\pi\tau} ~~~\text{for}~ \frac{t}{\tau}=\pm 1.
\ea
Clearly, $\frac{\partial}{\partial t}\rho_1^{1,1} \sim \frac{\partial}{\partial t}\mathcal{G}^{1,1} \approx 0$ far away from the QCP 
where we can expect adiabatic dynamics even without the application of any control. In contrast,  
$\frac{\partial}{\partial t}\rho_1^{1,1} \ll \frac{\partial}{\partial t}\mathcal{G}^{1,1}$ near the QCP, which necessitates the
introduction of control fields of the form (\ref{hcon})
in order to achieve local transitionless dynamics. Following the analysis presented above for the general case, one gets 
\ba
\Delta_y (t \approx \pm \tau) \approx \frac{\rm{ln} L}{4 \pi \tau \epsilon}.
\label{deltasc}
\ea
In deriving the above result we have replaced 
$\rm{Re}\left[\rho_1^{1,2}\right] = 0$ by $0 < \epsilon \ll 1$.
As explained before, this restricts $\Delta_y$ to a finite value, while having $\rho_1(t) \approx \mathcal{G}(t)$, as
quantified by the infidelity $\left(1 - \rm{tr}\sqrt{\sqrt{\mathcal{G}}\rho_1\sqrt{\mathcal{G}}} \right)$ (see Fig. (\ref{delta_vs_t}) and Appendix). 
Clearly, this conforms to our
general scaling relations for the transverse Ising model with $z = d = 1$ (see Eq. (\ref{scaly})). We note that keeping $\Delta_y (t \approx \pm \tau)$ fixed, scaling 
the system
size to $L^{'} = L^{x}$ results in 
\ba
\Delta_y (t \approx \pm \tau) = \frac{\rm{ln} L}{4 \pi \tau \epsilon} = \frac{\rm{ln} L^{'}}{4 \pi \tau x\epsilon},
\label{deltasc}
\ea
thus effectively linearly scaling the error to $x\epsilon$, thereby showing the robustness of our protocol.
Further, numerical analysis of the von-Neumann equation (\ref{vn}) for the single site reduced density matrix shows
$|J|, |h_z|$ increase with increasing $L$ at the QCP (see figs. (\ref{delta_vs_L}), (\ref{hJ}) and Appendix). 
It is worthwhile to note that the reduced fidelity susceptibility $\chi_1$ of a single spin in the 
transverse Ising model  scales as $\left(\rm{ln} L\right)^2$ at the QCP, thus once more verifying our general scaling relation (see Eq. (\ref{suscsc}))
\ba
\Delta_y  \sim \frac{\partial \lambda}{\partial t} \chi^{1/2} \sim \frac{\rm{ln} L}{\tau}. 
\ea


\section{Local transitionless dynamics in a spin system with long-range interactions}

\subsection{Central spin model: Generic case}
\label{STA2A}
In this section we remove the constraint of a single spin interacting with its nearest neighbors only; in contrast, we extend our analysis to a central spin (CS) model where a single central
spin $s$ interacts uniformly with all spins of an environmental spin chain $E$ (see Fig. (\ref{schem1}c)). The CS model has been investigated extensively in a series of works in the recent past in problems related to
decoherence, Lochsmidt echo, non-markovianity of baths to name a few \ct{zurek03, zurek11, rossini07, mukherjee12}. Such a system can be represented by the Hamiltonian
\ba
H^{cs}_{free} = H_0 + \lambda V_{\lambda} + r \ket{\uparrow}_s {_s}\bra{\uparrow}V_{\lambda} + \mathcal{B}^{z}_{free,s}(t)\sigma_s^z.
\label{Hcs}
\ea
Here $H_0$ is the critical Hamiltonian of the environmental spin chain and as before, $\lambda$ denotes the deviation from the QCP ($\lambda = 0$). $V_{\lambda}$ is an operator
acting on the environment.
The term $r \ket{\uparrow}_s {_s}\bra{\uparrow}V_{\lambda}$ is responsible for the global coupling between $s$ and $E$, where we have assumed the coupling exists only if 
$s$ is in the $\ket{\uparrow}_s$ state. The composite system starts from an initial product state at $t = -T$, with the spin $s$ being in its 
local ground state $\ket{s}(t=-T) = \ket{\uparrow}$ with respect to the time dependent local field $\mathcal{B}^{z}_{free,s}(t=-T) < 0$ and the environment spin chain is the ground state $\ket{E_G}$ of
the Hamiltonian (\ref{Hcs}) with $\mathcal{B}^{z}_{free,s} = 0$. 
We assume $\mathcal{B}^{z}_{free,s}(t)$ changes sign at $t = 0$, at which 
point $s$ stops being in its local ground state. Naturally, transitionless driving in this case would correspond to flipping the CS
to $\ket{s} = \ket{\downarrow}$ at $t = 0^{+}$, thereby keeping it in its instantaneous ground state at all times.
For a product initial state of the form
$\ket{\psi} = \left(c_+\ket{\uparrow} + c_-\ket{\downarrow}\right) \otimes \ket{E_G}$,  the reduced density matrix of the CS evolves as \ct{zurek11}
\ba
\rho_s(\delta t) = \left[\begin{array}{cc} |c_+|^2 & c_+ c_-^* f^*(\delta t) \\ c_+^* c_- f(\delta t) & |c_-|^2 \end{array} \right]
\label{LE}
\ea
for small time $\delta t$, where \ct{mukherjee12}
\ba
|f(\delta t)|^2 &\approx& exp\left[-\alpha (\delta t)^2 \right] \non\\
\alpha &\sim& r^2 L^{2(1/\nu - z)} ~~~ \text{if}~ \lambda \ll L^{-1/\nu}\non\\
&\sim& r^2 \lambda^{2(\nu z - 1)} ~~~~~\text{if}~ \lambda \ll L^{-1/\nu}
\ea
However, the above scaling relations are valid as long  $2/\nu - 2z > d$. Otherwise in higher dimensional systems the contribution coming from the low energy modes become subleading, and we 
get $\alpha \sim L^{d}$. The decay of $f$ is responsible for the decrease in purity of $s$ for $c_{+}, c_{-} \neq 0$. Clearly, one needs to
flip $s$ 
in a time $\delta t << 1/\sqrt{\alpha}$ to ensure that the spin does not lose its purity during the process. This requires the application of a control Hamiltonian 
 of the form $\Delta_{cs} \sigma_s^x$ at $t = 0$ such that 
\ba
|\Delta_{cs}| = \frac{\pi}{2\delta t} &\gg& \frac{\pi r L^{1/\nu - z}}{2}   ~~ \text{if}~  \lambda \ll L^{-1/\nu} \non\\
&\gg& \frac{\pi r\lambda^{\nu z - 1}}{2}   ~~ \text{if}~ \lambda \gg L^{-1/\nu}
\label{csgen}
\ea
We note that as mentioned above, in higher dimensions $d > 2/\nu - 2z $, the scaling forms (\ref{csgen}) reduce to $|\Delta_{cs}| \gg  L^{d/2}$. Clearly, $\Delta_{cs}$ is independent of the exact
time dependence of $ \mathcal{B}^{z}_{free,s}(t)$.
In the other limit of the CS interacting with a finite number of spins instead of the global coupling considered in Eq. (\ref{Hcs}), $\alpha$ does not scale with system size any more and
we simply have 
$|\Delta_{cs}| \gg \pi r/2$  \ct{rossini07}.

\subsection{Central spin model: Transverse Ising environment}
\label{STA2B}
Let us now focus on the specific example of the transverse Ising model in $d = 1$. Our composite Hamiltonian now takes the form 
\ba
H^{cs}_{free,I} &=& -\sum_j \sigma_j^x \sigma_{j+1}^x + (\lambda - 1)\sum_j \sigma_j^z \non\\ &+& r \ket{\uparrow}_s {_s}\bra{\uparrow}\sum_j \sigma_j^z  +  \mathcal{B}^{z}_{free,s}(t)\sigma_s^z,
\ea
In this case $\nu = z = d = 1$, and $\alpha$ scales linearly with $L$ \ct{mukherjee12}. Therefore proceeding as before for the generic case and demanding LSTA in the central spin, one arrives at the scaling relation
\ba
|\Delta_{cs}| \gg \frac{\pi r L^{1/2}}{2}.
\ea
The above relation agrees with the general scaling relations presented in section (\ref{STA2A}) and once again shows the possibility of achieving LSTA even in presence of non-local interactions. Interestingly, we reach this goal 
by application of control fields which scale  sub-extensively with 
system size.

Following some related  works which  have focused on central spin model with two central spins interacting with a spin chain environment, we now extend 
our analysis to two central spins $s1$ and $s2$, described by the Hamiltonian \ct{yuan07}
\ba
H^{cs}_{free,I2} &=&  -\sum_{j=1}^L \sigma_j^x \sigma_{j+1}^x + (\lambda - 1)\sum_{j=1}^L \sigma_j^z \non\\ &-&\frac{ r}{2}\left(\sigma_{s1}^z + \sigma_{s2}^z \right) \sum_{j=1}^L \sigma_j^z  +  \mathcal{B}^{z}_{free,s}(t)\left( \sigma_{s1}^z + \sigma_{s2}^z \right) \non.
\ea
As before, we assume $sgn( \mathcal{B}^{z}_{free,s}(t)) = sgn(t)$, the two central spins start in their local ground state $\ket{\uparrow \uparrow}$ at
time $t = t_{in}<0$, while the spin chain environment starts from an initial ground state corresponding to $H(r = h_s = 0)$. One can 
show that as before, the off-diagonal terms of the two central spin reduced density matrix decay at a rate $\exp\left[-\alpha \delta t^2 \right]$. Once again LSTA in this case for $g \to \pm 1$ 
would require flipping the central spins to $\ket{\downarrow \downarrow}$ state by application of local control fields of the form
$\Delta_{cs}\left(\sigma_{s1}^x + \sigma_{s2}^x\right)$ with $|\Delta_{cs}| \gg \pi r L^{1/2}/2$.

\section{Conclusion}
\label{concl}
In conclusion, we have shown the possibility of generating LSTA in quantum critical spin systems driven out of 
equilibrium by application of local 
control Hamiltonians. The subsystem under consideration can range from a single spin to half of the spin system. We have derived generic scaling forms for the control fields 
with the system size near a QCP.
Interestingly, the control fields can be related to reduced fidelity susceptibility as well.  We have also extended our analysis to transitionless dynamics in a central spin model with one or two
central spins, where the central spin(s) interact uniformly
with all spins of an environmental spin chain. 
 Our studies points to a lower bound of the control field in this case,
 which again scales with the system size following a generic scaling law.
  Finally we have verified our generic scaling results using the exactly solvable transverse Ising model in one 
 dimension, where the control fields scale sub-extensively with the system size, thus offering the possibility of engineering 
 LSTA with robust control protocols even in large systems.
 
 \acknowledgments
 We acknowledge financial support from the EU integrated 
projects SIQS, RYSQ and QUIC, from Italian MIUR via PRIN Project 2010LLKJBX, from the DFG via the SFB/TRR21 and from Ricerca SNS  R.F acknowledges  the Oxford Martin 
School  for support and the Clarendon Laboratory for hospitality during the completion of the work. V.M. also thanks A. Dutta and W. Niedenzu for helpful discussions. 
 
\section{Appendix}

We determine $h_z$ and $J$ by solving the set of coupled
differential equations near the QCP $t \approx \pm \tau$, obtained from Eq. (\ref{vn}) viz.
\ba
\frac{d}{dt} y_1(t) &=& 2h_z y_2(t) + \Delta_y\left[2\left(y_3(t) + y_4(t) \right) - 1 \right] \non\\
\frac{d}{dt} y_2(t) &=& -2h_z y_1(t) - 2J\epsilon \non\\ 
\frac{d}{dt} y_3(t) &=& \frac{I_G(t/\tau)}{4\pi\tau} - \Delta_y(t) y_1(t) \non\\
\frac{d}{dt} y_4(t) &=& -\frac{I_G(t/\tau)}{4\pi\tau} - \Delta_y(t) y_1(t) \non\\
\Delta_y (t) &=& \frac{I_G(t/\tau)}{4 \pi \tau \epsilon} \non\\
2J y_2(t) &=& \Delta_y (t) \left[\mathcal{G}^{1,1}(t) -  \mathcal{G}^{2,2}(t)\right]  \non\\
J y_1(t) &=& -h_z \epsilon,
\label{seven}
\ea
where $y_1 = \rm{Re}\left[\rho_2^{1,3} - \rho_2^{2,4}\right]$, $y_2 = \rm{Im}\left[\rho_2^{1,3} - \rho_2^{2,4}\right]$, $y_3 = \rho_2^{2,2}$ and $y_4 = \rho_2^{3,3}$ and 
\ba
I_G(x) = \int^{\pi}_0 \frac{sin^2 k}{\left(\cos k + x \right)^2 + \sin^2 k} dk.
\label{Igapp}
\ea

As initial condition we assume the system starts from its ground state at time $t = t_0 \to -\tau^{-}$. 

We note that $J(t)$ and hence $h_z(t)$ may diverge in the limit $y_2(t) \to 0$. In this case one can replace $y_2(t)$ by a 
second error parameter $ \rm{sign}(y_2(t))\epsilon_2$ (where as before, $0 < \epsilon_2 \ll 1$)
for $|y_2(t)| < \epsilon_2$ in order to restrict $J(t)$ and $h_z(t)$ to finite values. Here 
$\rm{sign}(y_2(t)) = 1 ~ (-1)$ for positive (negative) $y_2(t)$.

\end{document}